\documentclass[
    ,final            
  ]
  {aipproc}

\layoutstyle{8x11single}

\usepackage{graphicx}

\begin{document}

\title{Analytic Approach for Controlling Realistic Quantum Chaotic Systems}


\author{Toshiya~Takami}{
  address={Research Institute for Information Technology, Kyushu University,
Fukuoka 812--8581, Japan}
}

\author{Hiroshi~Fujisaki}{
  address={Institut f\"ur Physikalische und Theoretische Chemie,
J.~W.~Goethe-Universit\"at, Max-von-Laue-Str. 7,
D-60438 Frankfurt am Main, Germany}
}

\begin{abstract}
An analytic approach for controlling quantum states,
which was originally applied to fully random matrix systems
[T. Takami and H. Fujisaki, Phys. Rev. E {\bf 75}, 036219 (2007)],
is extended to deal with more realistic quantum systems
with a banded random matrix (BRM).
The validity of the new analytic field is confirmed by
directly solving the Schr\"odinger equation with a BRM interaction.
We find a threshold of the width of the BRM for the quantum control
to be successful.
\end{abstract}

\maketitle


\section{Introduction}

Theoretical and experimental studies of controlling quantum states
have been attracting much attention because of the theoretical
progress in the field of quantum computing \cite{NC00} and of the
technical developments in manipulating atomic and molecular
systems. Various approaches have been applied to such quantum systems
and its target is often "simple" such as one-state-to-one-state
control via adiabatic passage \cite{RZ00}.  When we consider to design
quantum devices with a large number of states interacting with a
complex environment or with short-time laser pulses, however, a
multi-state-to-multi-state control problem is not an exception but a
rule.  Optimal control theory \cite{RZ00} and genetic algorithms
\cite{RHR04,Miller06} are most successful methods to solve this kind
of complicated problem, but its implementation and interpretation can
be still difficult.

From theoretical points of view, a "complex" quantum system may be
modeled by fully random matrix systems with generic properties
\cite{Haake01} or by quantum chaos systems where their classical limit
shows chaotic properties \cite{Gutzwiller90}.  Gong and Brumer applied
the coherent control method \cite{SB86,SB03} to a quantum chaos system
\cite{GB01-PRL,GB05} and its prediction has been recently confirmed by
experiment.  We derived an analytic optimal field to control the fully
random matrix systems \cite{TF07} and the results are promising.
However, the most realistic quantum systems may be modeled with a
banded random matrix \cite{Haake01} and our previous method can not be
directly applied to such a more general situation.

In this contribution, we improve our previous approach \cite{TF07} to
deal with more realistic quantum systems with a banded random
Hamiltonian.  A new analytic field for optimal control is introduced and its
validity is evaluated by numerically solving the Schr\"odinger equation
for the multi-state-to-multi-state control problem.

\section{Analytic External Field for Controlling Quantum States\label{review}}

In the previous works \cite{TF07,TFM05,TF04},
we studied quantum control dynamics for a Hamiltonian
driven by an external field $\varepsilon(t)$,
\begin{equation}
\label{eqn:Hamiltonian}
H[\varepsilon(t)]=H_0+\varepsilon(t)V
\end{equation}
where $H_0$ and $V$ are fully random matrices.
For initial and target states in the eigenstate representation of $H_0$,
\begin{equation}
\label{eqn:initial_target}
|\Phi_0\rangle=\sum c_j|\varphi_j\rangle,\qquad
|\Phi_T\rangle=\sum d_k|\varphi_k\rangle
\end{equation}
an analytic optimal field was derived as 
\begin{equation}
\label{eqn:field}
  \varepsilon(t)=\frac{\pi\hbar}{T\left|V\right|^2}
  \sum_j\sum_k{\rm Re}\left[
    c_j^*V_{jk}d_k\ e^{iw_{jk}t}
  \right],\qquad
  V_{jk}=\langle\varphi_j|V|\varphi_k\rangle,\qquad
  w_{jk}=\frac{E_j-E_k}{\hbar}
\end{equation}
where $V_{jk}$ are elements for a fully random matrix
with $|V|^2\equiv\langle|V_{jk}|^2\rangle$.
Here $\langle \dot \rangle$ denotes an ensemble average.
Driven by this field, the quantum state $|\psi(t)\rangle$ is
shown to be steered from an initial state $|\Phi_0\rangle$ at $t=0$ to
a target state $|\Phi_T\rangle$ at $t=T$ according to
\begin{equation}
  |\psi(t)\rangle
  =\sum_ja_j(t)|\varphi_j\rangle e^{iw_{jk}t}
  =\hat U_0(t,0)|\Phi_0\rangle\cos\left(\frac{\pi t}{2T}\right)
    -i\hat U_0(t,T)|\Phi_T\rangle\sin\left(\frac{\pi t}{2T}\right),
\end{equation}
where $\hat U_0(t_2,t_1)$ is a propagator generated by $H_0$.

We can summerize the reason why the analytic field works well
for fully random matrix systems:
\begin{itemize}
\item The random phase property of $c_j$ and $d_j$ is used,
where the initial and target states are represented in linear combinations
of many eigenstates $|\varphi_j\rangle$.
If we introduce an approximate number of states,
$N\sim\sum|c_j|\sim\sum|d_k|$,
sums of complex numbers, $\sum{c_j}^2$, $\sum{d_j}^2$,
$\sum c_j^*d_j$, etc., are quantities with an order $O(1/N)$,
while $\sum|c_j|^2=\sum|d_j|^2=1$ from the normalization condition.
Thus, quantities of $O(1/N)$ can be ignored
compared to those of $O(1)$ for $N\rightarrow\infty$.
\item The rotating wave approximation is applicable, which is valid
when the control field amplitude is small enough.
This situation is satisfied when $T\rightarrow\infty$.
\end{itemize}
Using these properties, we proved the validity of our
analytic control field for fully random matrix systems \cite{TF07,TFM05,TF04}.

\section{Extension for Banded Random Matrix Systems}

We consider the case that the interaction Hamiltonian $V$ is
a banded random matrix in the eigenstate representation of $H_0$.
The elements of $V$ are random complex numbers with distribution
\begin{equation}
\label{eqn:V-brm}
  \left\langle\left|V_{jk}\right|^2\right\rangle
  \equiv\left\langle\left|\langle\phi_j|V|\phi_k\rangle\right|^2\right\rangle
  =\exp\left[-\frac{(E_j-E_k)^2}{{\Delta_0}^2}\right].
\end{equation}
We introduce an analytic optimal field as an extension of the analytic
field for fully random matrix systems,
\begin{equation}
  \varepsilon(t)=\sum_{jk}{\rm Re}\left[
    A_{jk}c_j^*V_{jk}d_k\ e^{iw_{jk}t}
  \right],
\end{equation}
with an extra-amplitude factor $A_{jk}$.
The coefficients $a_j(t)$ satisfy the Schr\"odinger equation
\begin{equation}
\label{eqn:rwa-cg}
  i\hbar\frac{d}{dt}a_k(t)=\frac12\sum_j\left[
    A_{jk}{c_j}^*d_k+{A_{kj}}^*c_k{d_j}^*
  \right]\left|V_{jk}\right|^2a_j(t)
\end{equation}
under the rotating-wave approximation.
If we assume that the transition is smooth,
$a_j(t)$ should be written as 
\begin{equation}
  a_k(t)=c_k\cos\left(\frac{\pi t}{2T}\right)
        -id_k\sin\left(\frac{\pi t}{2T}\right).
\end{equation}
Substituting these coefficients into Eq. (\ref{eqn:rwa-cg}),
we obtain a relation
\begin{equation}
  \frac{i\pi\hbar}{2T}\left[
    -c_k\sin\left(\frac{\pi t}{2T}\right)
    -id_k\cos\left(\frac{\pi t}{2T}\right)
  \right]=\frac12\sum_j\left[
      A_{jk}\left|c_j\right|^2d_k\cos\left(\frac{\pi t}{2T}\right)
     -iA_{kj}^*c_k\left|d_j\right|^2\sin\left(\frac{\pi t}{2T}\right)
  \right]\left|V_{jk}\right|^2.
\end{equation}
under the assumption of random phases.
Finally, we obtain conditions for the coefficients $A_{jk}$
\begin{equation}
  A_{jk}=A_{kj}^*\simeq\frac{\pi\hbar}{T}\exp\left[
    \frac{(E_j-E_k)^2}{{\Delta_0}^2}
  \right].
\end{equation}
If we consider the case that
those coefficients $c_j$ and $d_k$ have Gaussian distribution functions
in the energy space,
\begin{equation}
\label{eqn:initial_target-brm}
  \langle|c_j|^2\rangle
    \propto\exp\left[-\frac{(E_j-E_c)^2}{{\Delta_c}^2}\right],\qquad
  \langle|d_k|^2\rangle
    \propto\exp\left[-\frac{(E_k-E_d)^2}{{\Delta_d}^2}\right],
\end{equation}
with centers $E_c$ and $E_d$ and widths $\Delta_c$ and $\Delta_d$,
we can define the analytic optimal field as
\begin{equation}
\label{eqn:field-brm}
  \varepsilon(t)=\frac{\pi\hbar}{T}\sum_{jk}{\rm Re}\left[
    c_j^*V_{jk}d_k\ e^{iw_{jk}t}
  \right]\exp\left[\frac{(E_j-E_k)^2}{{\Delta_0}^2}\right].
\end{equation}
This field has a finite amplitude only when 
\begin{equation}
\Delta_c<\Delta_0\quad\hbox{and}\quad\Delta_d<\Delta_0.
\end{equation}
If not, the field has an infinite amplitude
in the limit of $E_j, E_k\rightarrow\pm\infty$
by the exponential factor $A_{jk}$ in Eq. (\ref{eqn:field-brm}).
Thus, the analytic field is refined
when the widths of the initial and target states are relatively small
compared to the width of the banded random matrix elements.

\subsection{Numerical evaluation}

\begin{figure}
\includegraphics[scale=0.6]{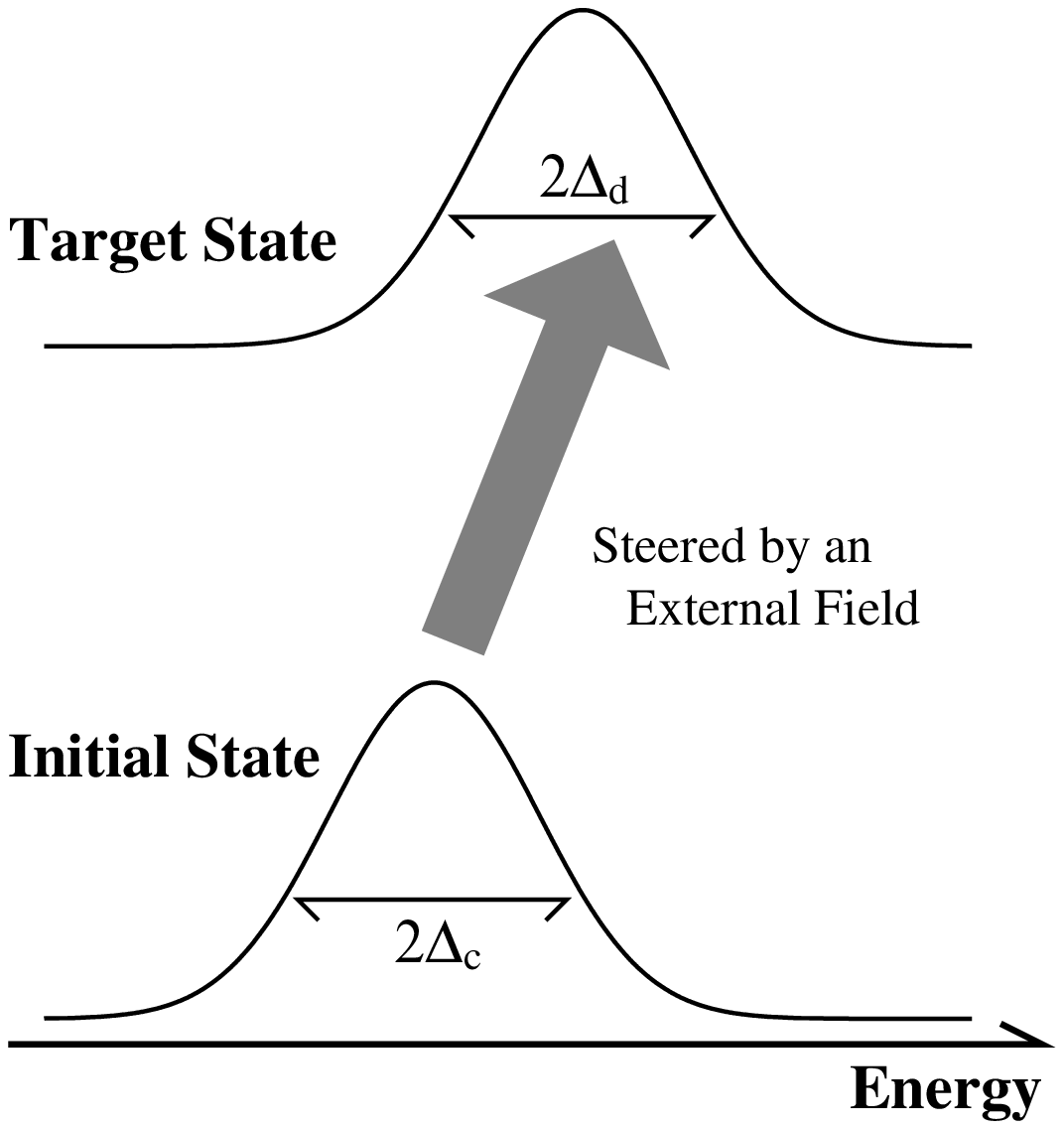}
\includegraphics[scale=0.4]{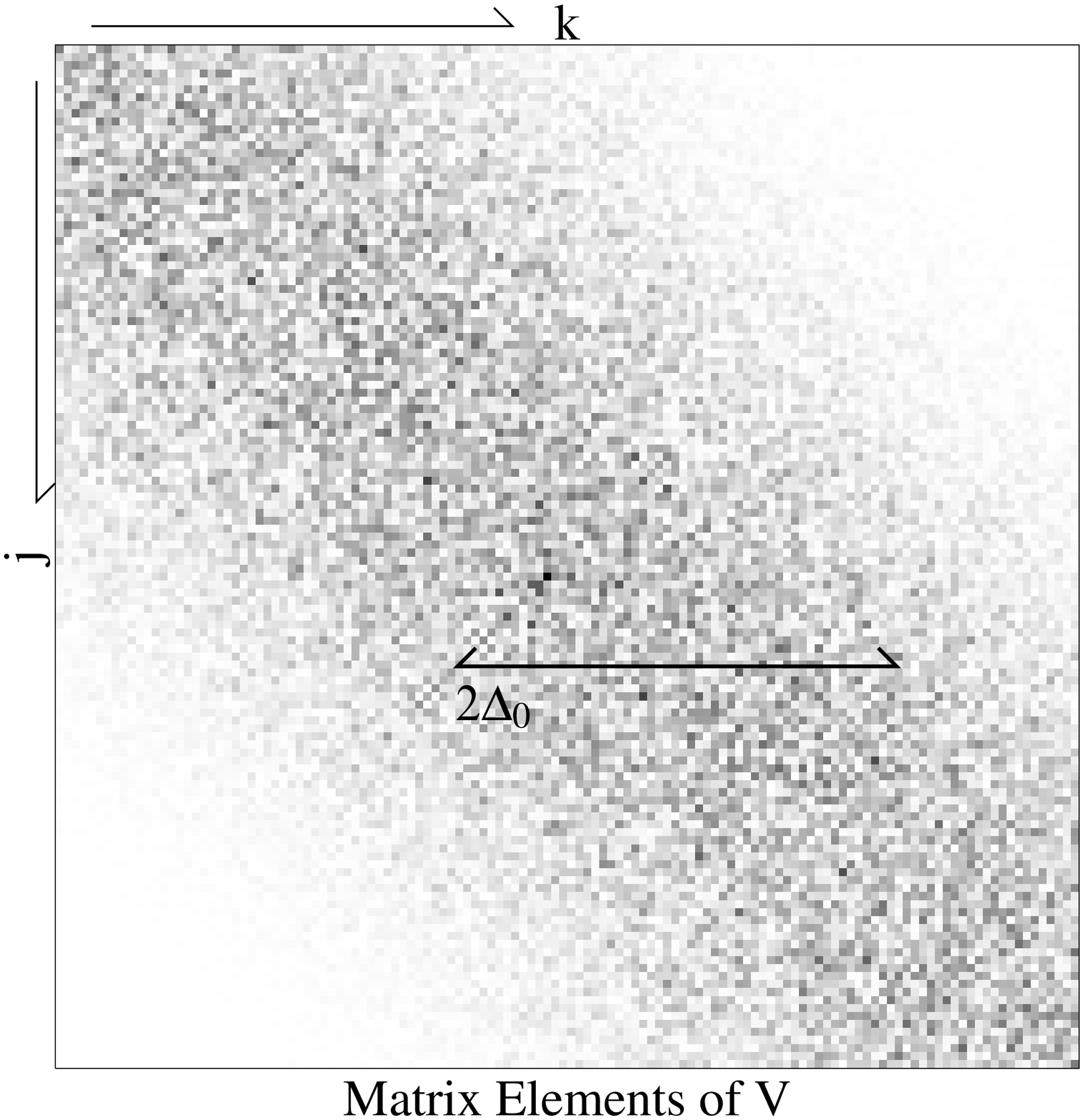}
\caption{
\label{fig:schematic}
Left: Schematic picture of the multi-state-to-multi-state control process. 
Right: An example of the interaction
Hamiltonian V ($128\times128$, $\Delta_0=32$), 
which is assumed to be a banded random matrix in 
the energy representation of $H_0$. 
}
\end{figure}

We shall confirm the validity of the new optimal field 
for the system with a banded random-matrix interaction.
The numerical test is configured as follows
(see Figure \ref{fig:schematic}).
The initial and target states are defined
as quantum vectors (\ref{eqn:initial_target})
with random complex coefficients $c_j$ and $d_j$
subject to (\ref{eqn:initial_target-brm}).
Here, we choose $\Delta_c=\Delta_d=32$, $E_c=-16$, and $E_d=16$,
where $H_0$ is a $128\times128$ random matrix
of the Gaussian orthogonal ensemble and is scaled so that
its eigenvalues $\{E_j\}$ are distributed in an interval $[-64,\ 64]$.
The interaction Hamiltonian $V$ is also a $128\times128$ matrix
while its elements obey a banded-random distribution (\ref{eqn:V-brm})
in the eigenstate representation of $H_0$ with $\Delta_0=32$.

\begin{figure}
\includegraphics[scale=0.6]{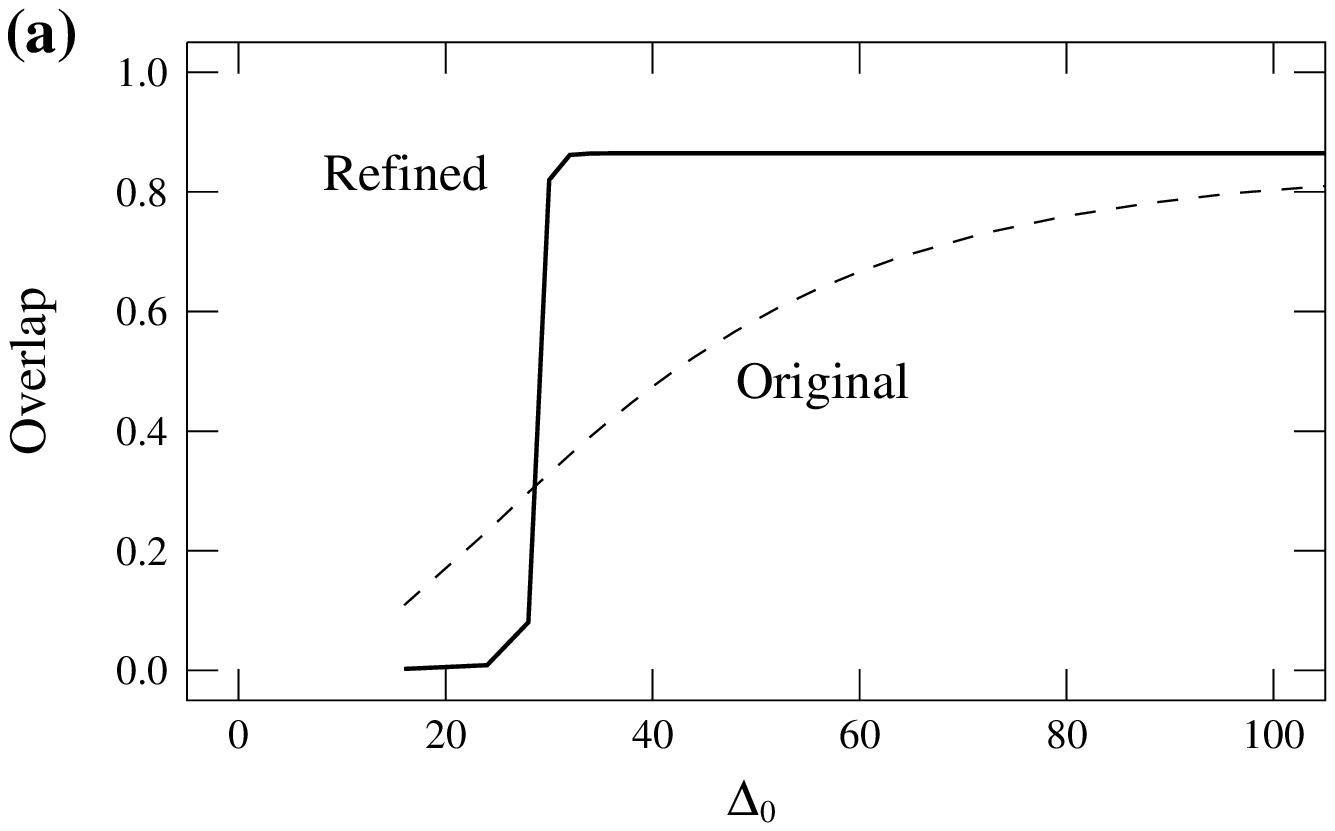}
\includegraphics[scale=0.6]{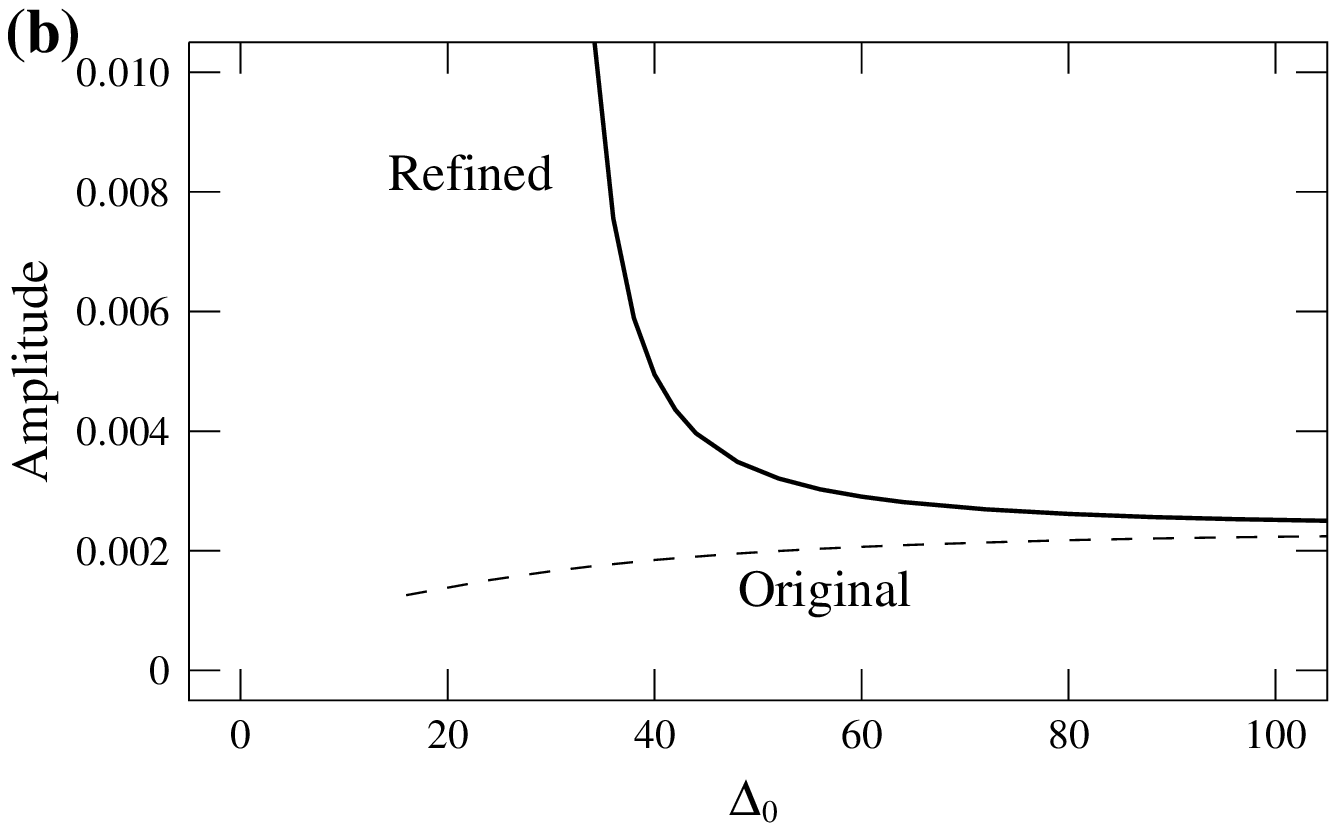}
\caption{
Numerical results for the multi-state-to-multi-state control problem.
The solid (dashed) curve 
represents the result by the new optimal field Eq. (\ref{eqn:field-brm}) 
(by the previous optimal field with $A_{jk}=1$).
(a) The final overlap is shown as a function of $\Delta_0$ (width of V).
(b) The amplitude (defined as a time average of the absolute square of a field) of 
the optimal field as a function of $\Delta_0$. 
}
\label{fig:result}
\end{figure}

The optimal field (\ref{eqn:field-brm}) is calculated
from those quantities $\{c_j\}$, $\{d_k\}$, $\{V_{jk} \}$, and $\{E_j\}$
with parameters $T$ and $\Delta_0$.
In order to check the validity of our optimal field (\ref{eqn:field-brm}),
we solve the initial value problems with Hamiltonian (\ref{eqn:Hamiltonian})
driven by the optimal field (\ref{eqn:field-brm})
for various band widths $\Delta_0$ of the interaction Hamiltonian $V$.
The results are shown in Figure \ref{fig:result}.
When we use the original analytic field (\ref{eqn:field}),
the performance of the optimal field (dashed curve) decreases
for the banded matrices with smaller widths.
On the other hand, the final overlaps (solid curve) by the refined
analytic field (\ref{eqn:field-brm}) does not change
even for the smaller width
untill the limit $\Delta_0\approx\Delta_c$ or $\Delta_0\approx\Delta_d$.

%

%

\section{Conclusion}

We extended our previous analytic approach for controlling complex
quantum systems to deal with more realistic systems with a banded
random matrix.  The key ingredient is the amplitude factor $A_{jk}$,
which is an exponentially growing function, introduced in the analytic 
optimal field Eq. (\ref{eqn:field-brm}).  We showed that the new 
analytic optimal field outperforms the previous optimal field for the
multi-state-to-multi-state control problem.  Interestingly we found a
threshold of the width of the banded random Hamiltonian $\Delta_0$ for
the control to be successful: $\Delta_0\approx\Delta_c,\ \Delta_d$
where the latters are the width of the energy spreading of the initial
and final states.  In the near future we will apply this optimal field
to quantum chaos systems such as quantum kicked rotors (tops) and to
more realistic molecular systems \cite{TF07-p}.



\begin{theacknowledgments}
We are grateful to the Research Institute for Information Technology
(RIIT) in Kyushu University for providing the computer resources for
the numerical calculations presented here.
\end{theacknowledgments}



\bibliographystyle{aipproc}   

\bibliography{BibTeX/MyWorks,BibTeX/Control,BibTeX/QuantumChaos}


\end{document}